\documentclass[prb,preprint,aps]{revtex4} 
\usepackage{graphicx} 
\usepackage{tabularx}
\usepackage{dcolumn} 
\usepackage{color}
\usepackage{bm}
\usepackage{amsmath}
\usepackage{amssymb}

\begin{document} 
 
\title{%
Microscopic Eilenberger theory of Fulde-Ferrell-Larkin-Ovchinnikov states in the presence of vortices
}

\author{Kenta M. Suzuki} 
\affiliation{Department of Physics, Okayama University, Okayama 700-8530, Japan} 

\author{Kazushige Machida} 
\affiliation{Department of Physics, Ritsumeikan University, 
Kusatsu 525-8577, Japan} 

\author{Yasumasa Tsutsumi} 
\affiliation{RIKEN Center for Emergent Matter Science, Wako, Saitama 351-0198, Japan}

\author{Masanori Ichioka} 
\affiliation{Department of Physics, Okayama University, Okayama 700-8530, Japan} 
\affiliation{Research Institute for Interdisciplinary Science, Okayama University, Okayama 700-8530, Japan}
%\date{\today} 

\begin{abstract}
We theoretically investigate the Fulde-Ferrell-Larkin-Ovchinnikov (FFLO) state
by using the microscopic quasi-classical Eilenberger equation.
The Pauli paramagnetic effects and the orbital depairing effects due to vortices
are treated in an equal footing for three dimensional spherical Fermi surface model and $s$-wave pairing. 
The field evolution of the LO state is studied in detail, such as 
the $H$-$T$ phase diagram, spatial structures of the order parameter, the paramagnetic moment, and the
internal filed. Field-dependences of various thermodynamic quantities: 
the paramagnetic moment, entropy, and the zero-energy density of states
are calculated.  Those quantities are shown to start quickly growing upon entering the LO state.
We also evaluate the wave length of the LO modulation, the flux line lattice form factors for small angle neutron scattering,
and the NMR spectra to facilitate the identification of the LO state. Two cases of strong and intermediate Pauli paramagnetic effect
are studied comparatively. 
The possibility of the LO phase in Sr$_2$RuO$_4$, CeCoIn$_5$, CeCu$_2$Si$_2$,
and the organic superconductors is critically examined and crucial experiments to identify it
are proposed.
\end{abstract}
 
\maketitle 

%===introduction
\section{Introduction}

Fulde and Ferrell (FF)~\cite{FF}, and Larkin and Ovchinnikov (LO)~\cite{LO} in 1964 
proposed a theoretical possibility of spatially modulated 
superconducting state~\cite{matsuda} under Zeeman effect.
Since then, there have been a lot of works focusing on the realization of the FFLO state
both theoretically and experimentally. Yet there is no well-accepted material forz
the FFLO state.

In the FFLO state, the superconducting order parameter in the singlet pairing, 
such as $s$-wave or $d$-wave pairing, exhibits a spatial modulation\cite{matsuda}.
Under the population imbalance of up and down spin species of Cooper pairs, 
it is expected that FFLO is most possible state to emerge\cite{mizushima,machida}.
The population imbalance is brought about either by its preparation in cold neutral atom
gases \cite{zwier1,zwier2,randy} or by application of an external field in charged particle case through
the Pauli paramagnetic effect.

A part of the reasons of difficulties to realize the FFLO in a superconductor
may come from lack of theoretical investigations which
fully take into account both Pauli paramagnetic effect and 
flux line effect on an equal footing.
The simultaneous consideration of the two depairing effects; paramagnetic depairing in the 
former and the orbital depairing in the latter is a difficult task 
because the two kinds of spatial modulations, one is
due to the FFLO and the other is flux line lattice, must be handled simultaneously.
It is often the cases~\cite{takada,nakanishi,matsuo,shimahara,sauls} only to consider the Pauli paramagnetic effect by neglecting the latter effect,
including the original works by Fulde and Ferrell~\cite{FF}, and Larkin and Ovchinnikov~\cite{LO}.
In those studies $s$-wave~\cite{takada,nakanishi}  and $d$-wave~\cite{matsuo,shimahara,sauls}  pairing cases are treated.
 The attempts to simultaneously consider the two effects are limited to the so-called Ginzburg-Landau (GL) region
near $H_{\rm c2}$~\cite{gunther,tachiki,ikeda}. Thus we need more extensive studies which cover the whole region of $T$ and
$H$. This is one of our main purposes of the present paper.

The Larkin-Ovchinnikov (LO) state with periodically modulated amplitude of the order parameter 
is far more difficult to describe due to the so-called solitonic spatial variation with infinitely 
many higher harmonics of the Fourier component of the order parameter 
in general. This is handled exactly and analytically~\cite{nakanishi} 
only in the absence of the orbital depairing.
The LO state is so computationally demanding, but it
 is stabler than  the Fulde-Ferrell (FF) state 
where only the phase is modulated in the order parameter~\cite{nakanishi}. 
Thus we consider the LO state in this paper.
There are two possible modulation directions with respect to the applied magnetic field:
longitudinal and  transverse. In this paper, we consider the longitudinal LO state
which is expected to be stabler than the transverse LO state physically.

%Recently two important experiments; 
%nuclear magnetic resonance (NMR)~\cite{curro,vesna,kumagai1,kumagai2} and  
%neutron scattering~\cite{kenzelmann,morten} 
%%have been done and are still ongoing on a heavy Fermion superconductor
%CeCoIn$_5$ whose high field ($H$) and low temperature ($T$) phase is regarded  
%to be a prime candidate for FFLO state.
%It will turn out that these two methods can give direct and crucial evidence for FFLO,
%together with other variety of experiments\cite{bianchi}. 

Thus the main purpose of this paper is to provide fundamental 
theoretical information on the physical properties of the LO states.
In particular, we study how the field evolutions of various observables are, including thermodynamic quantities, such as 
the entropy, the zero energy density of states (DOS) measured by low temperature specific heat experiment,
and magnetization changes. We also calculate the flux line lattice (FLL)
form factors measured by small angle neutron scattering (SANS), and 
the nuclear magnetic resonance (NMR) spectrum in the LO state.

For that  purpose, to obtain the magnetic field $H$-dependence of the LO states
by advancing our previous study~\cite{ichiokaFFLO},   
we solve the microscopic Eilenberger equation 
fully selfconsistently in three-dimensional (3D) space of vortex and 
LO modulation~\cite{ichiokaFFLO}, and find free energy minimum with respect to the LO period $L$. 
The orbital depairing and Pauli paramagnetic depairing 
are treated in an equal footing here.
The phase diagram in $H$-$T$ plane is constructed where the Abrikosov phase
and LO phase are competing, and
we examine the behaviors of various observables mentioned above.
In this paper, we compare two cases of strong and intermediate Pauli paramagnetic effect. 

Our basic strategy is to study the canonical field-dependent properties 
of the LO states for spherical Fermi surface model and $s$-wave pairing.
The corresponding 3D calculation for the FF state~\cite{takada} 
and full selfconsistent analytic theory 
for quasi-1D case\cite{nakanishi} have been performed 
before without vortices.
Here we extend their calculations to take account of vortex effects. 
The effects of the $d$-wave pairing on the LO within the same Eilenberger framework
were reported~\cite{ichiokaFFLO}.

There are several important and outstanding experimental results to suggest the
LO state which remains unexplored in detail because of the lack of appropriate theoretical 
methods to describe further detailed behaviors of the LO state. 
For example, 

(1) NMR experiments on CeCoIn$_5$  where the resonance 
spectra exhibit a characteristic signature and change when entering 
the high field LO state~\cite{curro,kumagai1,kumagai2}.
And specific heat studies on  CeCoIn$_5$ for $H\parallel ab$ exhibit
a characteristic first order transition~\cite{bianchi}.
Neutron experiments~\cite{kenzelmann} detect anomalous magnetism, so-called Q phase
 in high field region for $H\parallel ab$.

(2) Small angle neutron scattering experiments have been done for CeCoIn$_5$ of  $H\parallel c$~\cite{morten,white}
where the FLL form factor $|F_{100}(H)|$ increases toward H$_{\rm c2}$ contrary to the ordinary type II superconductors which
exhibits a rapid decrease as $H$ increases. Just before $H_{\rm c2}$,  $|F_{100}(H)|$ sharply drops to zero.

(3) In $\kappa$-(BEDT-TTF)$_2$Cu(NCS)$_2$, 
Mayaffre {\it et al.}~\cite{vesna1} find a sharp increase of $T_1^{-1}$ as a function of $H$ and $T$ near and just below $H_{\rm c2}$
when entering the high field phase, suggesting the LO state in this quasi-2D superconductor.

(4) However, in CeCu$_2$Si$_2$~\cite{kitagawa} a similar $T_1^{-1}$ enhancement phenomenon is reported.
In our opinion it is unrelated to the LO although the authors claim it
because of the reasons given in Ref.~\onlinecite{CeCu2Si2}.
Thus  it is obvious that we definitely need a careful theoretical study to firmly identify
the LO, which is able to check various aspects of the LO signatures, not only
single phenomenon such as the $T_1^{-1}$ enhancement, but also the consistency with other phenomena
associated with the LO to avoid further confusion.

(5) Sr$_2$RuO$_4$, which was a prime candidate of  a chiral $p$ wave superconductor~\cite{maeno1,maeno2}, 
but recent various theoretical and experimental studies~\cite{irie,kittakaM,stuart} indicate now 
that it is most likely to be a spin singlet superconductor.
Since the system well satisfies the necessary conditions for the LO to appear.
Namely, it is super-clean in that (A) the mean free path $l$ must be longer the periodicity $L$ of the LO,
which is typically an order of 100$\xi$ with $\xi$ coherence length (see later for details), that is, $l \gg L$.
(B) Favorably, it is low-dimensional, and (C) strong Pauli paramagnetic effect to avoid the orbital depairing.
Thus it is enough reasons to investigate the LO in this material, which is true for other materials,
CeCoIn$_5$ and $\kappa$-(BEDT-TTF)$_2$Cu(NCS)$_2$, but no for CeCu$_2$Si$_2$
which is known to be a barely clean system, namely $l\sim \xi$ and three dimensional electronic structure
although the Pauli paramagnetic effect is sufficiently strong~\cite{CeCu2Si2}.

The plan of this paper is as follows.
We first introduce our formulation based on the microscopic quasi-classical Eilenberger framework~\cite{eilen} in Section II.
This formulation is valid for $\xi k_{\rm F}\gg 1$ with $k_{\rm F}$ the Fermi wave number, which is 
well satisfied for the materials of interest. The LO phase diagram in the $H$ vs $T$ plane is determined in Section III.
The spatial structure of the LO state is examined in Section IV.
The field evolutions of thermodynamic quantities mentioned above are presented in Section V. Those are accessible by
a variety of experimental methods. The FLL form factors with various indices and NMR spectra are calculated in Section VI
and VII respectively.
Throughout this paper, we treat two cases $\mu$=5 and $\mu$=2 comparatively, 
corresponding to strong and intermediate Pauli paramagnetic effect cases where $\mu$ is a measure of the
strength of the Zeeman effect, related to the so-called Maki parameter $\alpha_{\rm Maki}$ through
$\mu=2\alpha_{\rm Maki}$. 
In Section VIII, we examine critically each candidate material for the possible realization of the LO
state in light of the present calculation and propose further experiments to firmly establish and
identify the LO. We devote to conclusions in the last section.
A part of the present results is reported in Ref.~\onlinecite{kentasuzuki}.

\section{Formulation for Eilenberger theory}

%%%%%%%%%%%%%%%%%%%%%%%%%%%%%%%%%%%%%%%%%%%%%%%%%%%%%%%%%%%%%%%% 
%\section{formulation and notations} 

We calculate the 3D spatial structure of the vortex lattice state
by quasiclassical Eilenberger theory in the clean 
limit~\cite{ichiokaS,miranovic,hasegawa1,hasegawa2},
assuming that the order parameter modulates along the magnetic field direction
in the LO state.
The Pauli paramagnetic effects are included 
through the Zeeman term $\mu_{\rm B}B({\bf r})$, 
where $B({\bf r})$ is the flux density of the internal field and 
$\mu_{\rm B}$ is a renormalized Bohr 
magneton.
The quasiclassical Green's functions
$g( \omega_n +{\rm i} {\mu} B, {\bf k},{\bf r})$, 
$f( \omega_n +{\rm i} {\mu} B, {\bf k},{\bf r})$, and 
$f^\dagger( \omega_n +{\rm i} {\mu} B, {\bf k},{\bf r})$  
are calculated in the vortex lattice state  
by the Eilenberger equations~\cite{ichiokaFFLO,ichiokaS,miranovic,hasegawa1,hasegawa2,ichioka} 
\begin{eqnarray} &&
\left\{ \omega_n +{\rm i}{\mu}B 
+\tilde{\bf v} \cdot\left(\nabla+{\rm i}{\bf A} \right)\right\} f
=\Delta g, 
\nonumber 
%\label{eq:eil1}
\\ && 
\left\{ \omega_n +{\rm i}{\mu}B 
-\tilde{\bf v} \cdot\left( \nabla-{\rm i}{\bf A} \right)\right\} f^\dagger
=\Delta^\ast g  , \quad 
%\label{eq:eil2}
\label{eq:Eil}
\end{eqnarray} 
where $g=(1-ff^\dagger)^{1/2}$, ${\rm Re} g > 0$, 
$\tilde{\bf v}={\bf v}/v_{{\rm F}0}$, 
and the Pauli parameter ${\mu}=\mu_{\rm B} B_0/\pi k_{\rm B}T_{\rm c}$. 
${\bf k}$
%=(k_a,k_b,k_c)$
 is the relative momentum of the Cooper pair, 
and ${\bf r}$ is the center-of-mass coordinate of the pair. 
${\bf v}$ is the Fermi velocity and
$v_{\rm F0}=\langle v^2 \rangle_{\bf k}^{1/2}$ 
%is the averaged Fermi velocity on the Fermi surface 
where $\langle \cdots \rangle_{\bf k}$ indicates the Fermi surface average. 
Isotropic spherical Fermi surface is considered in this study. 
We assume that a magnetic field is applied to the z-axis.
The Eilenberger units $R_0$ for lengths and $B_0$ for magnetic fields 
are used\cite{ichioka,ichiokaFFLO}.
%The energy $E$, 
The order parameter $\Delta$ and the Matsubara frequency $\omega_n$ 
are normalized in units of $\pi k_{\rm B} T_{\rm c}$.

As for selfconsistent conditions, 
the order parameter is calculated by 
\begin{eqnarray}
\Delta({\bf r})
= g_0N_0 T \sum_{0 < \omega_n \le \omega_{\rm cut}} 
 \left\langle %\phi^\ast({\bf k}) \left( 
    f +{f^\dagger}^\ast %\right) 
\right\rangle_{\bf k} 
\label{eq:scD} 
\end{eqnarray} 
\noindent
with 
$(g_0N_0)^{-1}=  \ln T +2 T
        \sum_{0 < \omega_n \le \omega_{\rm cut}}\omega_n^{-1} $. 
We use $\omega_{\rm cut}=20 k_{\rm B}T_{\rm c}$.
%The vector potential for 
${\bf B}=\nabla\times{\bf A}$ 
with the vector potential 
${\bf A}=\frac{1}{2}\bar{\bf B}\times{\bf r} + {\bf a}$ 
and $\bar{\bf B}=(0,0,\bar{B})$. 
$\bar{B}$ is the averaged flux density of the internal field, 
and $\langle \nabla\times{\bf a}\rangle_{\bf r}=0$. 
The spatial variation of the internal field $\nabla\times{\bf a}$ 
is selfconsistently determined by 
\begin{eqnarray}
\nabla\times \left( \nabla \times {\bf a} \right) 
=\nabla\times {\bf M}_{\rm para}({\bf r})
-\frac{2T}{{{\kappa}}^2}  \sum_{0 < \omega_n} 
 \left\langle \tilde{\bf v} 
         {\rm Im}~g  
 \right\rangle_{\bf k}, 
\label{eq:scH} 
\end{eqnarray} 
where we consider both the diamagnetic contribution of 
supercurrent in the last term 
and the contribution of the paramagnetic moment 
${\bf M}_{\rm para}({\bf r})=(0,0,M_{\rm para}({\bf r}))$ 
with 
\begin{eqnarray}
M_{\rm para}({\bf r})
=M_0 \left( 
\frac{B({\bf r})}{\bar{B}} 
- \frac{2T}{{\mu} \bar{B} }  
\sum_{0 < \omega_n}  \left\langle {\rm Im}~g 
 \right\rangle_{\bf k}
\right) . 
\label{eq:scM} 
\end{eqnarray} 
The normal state paramagnetic moment 
$M_0 = ({{\mu}}/{{\kappa}})^2 \bar{B} $,   
${\kappa}=B_0/\pi k_{\rm B}T_{\rm c}\sqrt{8\pi N_0}$  and 
$N_0$ is the DOS at the Fermi energy in the normal state. 
We set the GL parameter $\kappa=102$.
Using the spatial averaged value 
$M_{\rm para}=\langle M_{\rm para}({\bf r})\rangle_{\bf r}$,  
the normalized paramagnetic susceptibility is given by 
$\chi_{\rm spin}=M_{\rm para}/M_0$. 

In Eilenberger theory, the Gibbs free energy is given by~\cite{hiragi}  
 %\begin{eqnarray} &&  
% F=
%\int_{\rm unit cell} {\rm d}{\bf r} \Bigl\{ 
% {\kappa}^2 |{\bf B}({\bf r})-{\bf H}|^2 
%-{\mu}^2 |B({\bf r})|^2  
%\nonumber \\ && \qquad
 %+ |\Delta({\bf r})|^2 ( 
%   \ln T + 2 T \sum_{0<\omega_n<\omega_{\rm cut}} {1\over \omega_n} )
% \nonumber \\ && \qquad
% -T  \sum_{|\omega_n|<\omega_{\rm cut}}
%    \left\langle I({\bf r},{\bf k},\omega_n) \right\rangle 
% \Bigr\}  
%\label{eq:f1}
%  \end{eqnarray}
% with 
% \begin{eqnarray} &&  
% I({\bf r},{\bf k},\omega_n)
% =\Delta \phi f^\dagger + \Delta^\ast \phi^\ast f 
% \nonumber \\ && \qquad
% +(g -\frac{\omega_n}{|\omega_n|})
% \Bigl\{ \frac{1}{f}\left( \omega_n +{\rm i}{\mu}B
%       + \hat{\bf v}\cdot(\nabla+{\rm i}{\bf A}) \right)f
% \nonumber \\ && \qquad\qquad
% + \frac{1}{f^\dagger}\left( \omega_n +{\rm i}{\mu}B
%       + \hat{\bf v}\cdot(\nabla-{\rm i}{\bf A}) \right)f^\dagger 
% \Bigr\} 
% \label{eq:f2}
% \end{eqnarray}
% in our dimensionless unit.
% Using Eqs. (\ref{eq:Eil}) and (\ref{eq:scD}), we obtain 
%\begin{eqnarray} &&
%F=
% \int_{\rm unit cell} {\rm d}{\bf r} \Bigl\{ 
%\left\langle 
%{\kappa}^2 |{\bf B}({\bf r})-{\bf H}|^2 -{\mu}^2 |B({\bf r})|^2 
%\right\rangle_{\bf r} 
%\nonumber \\ && 
%+T  \sum_{|\omega_n|<\omega_{\rm cut}}
%\left\langle 
%  {\rm Re} \left\langle 
%\frac{g-1}{g+1}(\Delta \phi f^\dagger + \Delta^\ast \phi^\ast f )
% \right\rangle_{\bf k}  \right\rangle_{\bf r}  . 
%\qquad 
%\Bigr\}  .
%\label{eq:f3}
%\end{eqnarray} 

\begin{eqnarray} &&
F=
% \int_{\rm unit cell} {\rm d}{\bf r} \Bigl\{ 
\left\langle 
{\kappa}^2 |{\bf B}({\bf r})-{\bf H}|^2 -{\mu}^2 |B({\bf r})|^2 
\right\rangle_{\bf r} 
%\nonumber \\ && 
+T  \sum_{|\omega_n|<\omega_{\rm cut}}
\left\langle 
  {\rm Re} \left\langle 
\frac{g-1}{g+1}(\Delta \phi f^\dagger + \Delta^\ast \phi^\ast f )
 \right\rangle_{\bf k}  \right\rangle_{\bf r}. 
\qquad 
%\Bigr\}  .
\label{eq:f3}
\end{eqnarray} 
$\langle \cdots \rangle_{\bf r}$ indicates the spatial average 
within a unit cell of the vortex lattice. 
The entropy in the superconducting state,  
given by $S_{\rm s}(T)=S_{\rm n}(T)- \partial F/\partial T$,  
is obtained 
% from Eqs. (\ref{eq:f1}) and (\ref{eq:f2}) 
as~\cite{hiragi}    

\begin{eqnarray} && 
\frac{S_{\rm s}(T)}{S_{\rm n}(T_{\rm c})}
=T - \frac{3}{2} 
%\int_{\rm unit cell} {\rm d}{\bf r} \Bigl\{ 
% \nonumber \\ && 
\sum_{0< \omega_n < \omega_{\rm cut}}{\rm Re} \Big\langle \Big\langle  
g_0N_0 
(\Delta \phi f^\dagger +\Delta^\ast \phi^\ast f )
%/( \ln T + 2 T{\sum}'_{\omega_n} {1\over \omega_n}) 
%\nonumber \\ && 
%\qquad\qquad
-2 \frac{ \Delta \phi f^\dagger +\Delta^\ast \phi^\ast f}{g+1} 
%\nonumber \\ && 
-4 \omega_n ( g-1 )
\Big\rangle_{\bf k}
\Big\rangle_{\bf r}
\qquad
\label{eq:S}
\end{eqnarray}

\noindent
where 
%${\sum}'$ implies $0<\omega_n<\omega_{\rm cut}$, and
$S_n$ is the entropy in the normal state.

We obtain the relation of $\bar{B}$ and 
the external field $H$ as 
\begin{eqnarray} && 
H=\left(1-\frac{{\mu}^2}{{\kappa}^2}\right) 
\left(
\bar{B}
+\frac{1}{\bar{B}} 
\left\langle \left( B({\bf r})-\bar{B} \right)^2\right\rangle_{\bf r}
\right)
\nonumber \\ &&   
+\frac{T}{{\kappa}^2 \bar{B}} \sum_{0 < \omega_n} 
\Bigl\langle  
\Bigl\langle 
{\mu} B({\bf r}) {\rm Im}  \left\{ g \right\}  
+\frac{1}{2}{\rm Re}\left\{ 
\frac{(f^\dagger \Delta+f \Delta^\ast)g}{g+1} \right\} 
%\nonumber \\ &&   
%\hspace{1cm}
+\omega_l {\rm Re}\{ g-1 \} 
\Bigr\rangle_{\bf k} \Bigr\rangle_{\bf r} 
\label{eq:H}
\end{eqnarray} 
from Doria-Gubernatis-Rainer scaling~\cite{watanabekita,doria}.  
%where $\langle \cdots \rangle_{\bf r}$ indicates the spatial average. 
%We consider the case of large Ginzburg-Landau (GL) parameter 
%$\kappa_{\rm GL} \sim {\kappa}=130$ 
%and low temperature $T/T_{\rm c}=0.1$. 
%For the two-dimensional Fermi surface, 
%${\kappa}
%=( 7 \zeta(3) /8 )^{1/2}  \kappa_{\rm GL} 
%\sim \kappa_{\rm  GL}$.\cite{miranovic2003}   
In the parameters used in our calculation, $|\bar{B}-H| < 10^{-4} B_0$. 
The magnetization is calculated as $M=\bar{B}-H$, 
which includes the paramagnetic component $M_{\rm para}$ 
in addition to the diamagnetic contributions. 

In the selfconsistent Eilenberger theory, 
we solve Eq. (\ref{eq:Eil}) and Eqs. (\ref{eq:scD})-(\ref{eq:scM})
alternately, and obtain selfconsistent solutions 
of $\Delta({\bf r})$, ${\bf A}({\bf r})$, 
and quasiclassical Green's functions with $\omega_n$, 
as in previous works~\cite{ichioka,ichiokaFFLO}
under a given unit cell of the triangular vortex lattice. 
Using the selfconsistent solutions, we evaluate 
the free energy in Eq. (\ref{eq:f3}), the entropy in Eq. (\ref{eq:S}), and the external field in Eq. (\ref{eq:H}). 

For the LO state, 
$\Delta({\bf r})$ has periodic oscillation of the period $L$ along the $z$ axis of the vortex line, 
in addition to the vortex lattice structure in the $xy$ plane. 
The unit cell of the vortex lattice is given by 
$(x,y)=u_1({\bf r}_1-{\bf r}_2)+u_2 {\bf r}_2$ with $-0.5 \le u_i \le 0.5$ ($i=1,2)$. 
${\bf r}_1=(c_x,0,0)$ and ${\bf r}_2=(c_x/2,c_y,0)$ with 
$c_x c_y \bar{B}=\phi_0$ and the flux quantum $\phi_0$. 
As the unit cell size of vortex lattice is determined by $\bar{B}\sim H$, 
we can estimate $H$-dependence of the LO states in our calculation of the vortex lattice.  
We use $\mu=5$ and $\mu=2$
as representative cases of strong  and intermediate Pauli paramagnetic effect, respectively.

When we calculate the electronic state, 
we solve Eq. (\ref{eq:Eil}) with $i\omega_n \rightarrow  E+i\eta$. 
In the calculation we use $\Delta({\bf r})$, ${\bf A}({\bf r})$, and $B({\bf r})$ which are obtained from the above selfconsistent calculation. 
$\eta$ is an infinitesimal constant. 
From the quasiclassical Green's function of real energy $E$, 
the DOS is given by $N(E)=(N_{+1}(E)+N_{-1}(E))/2$ with 
\begin{eqnarray}
N_\sigma(E)
=N_0 
{\rm Re} \left\langle\left\langle
g(\omega_n+{\rm i}\sigma \mu B,{\bf k},{\bf r})|_{i\omega_n \rightarrow  E+i\eta } 
\right\rangle_{\bf k}
\right\rangle_{\bf r}
\end{eqnarray}
with $\sigma=+1$ ($-1$) for the up (down) spin component. 
We study the $H$-dependence of the Sommerfeld coefficient $\gamma(H)$ 
of the low temperature specific heat. 
This is given by the normalized zero-energy DOS as $\gamma(H)=N(E=0)/N_0$.

\section{Phase diagram}

Before studying the thermodynamic quantities in the LO state mentioned above, 
we evaluate the phase diagram of the LO state, and the stable LO period $L$ as a function of $H$. 
The Gibbs free energy in Eq.~\eqref{eq:f3} is calculated from selfconsistent solutions 
of Eq.~\eqref{eq:scD} 
for the LO states with various LO wave length $L$ normalized by $R_0$. 
We compare them to find the most stable state under a given $H$ and $T$. 

\begin{figure}
\centering 
\includegraphics[width=14cm]{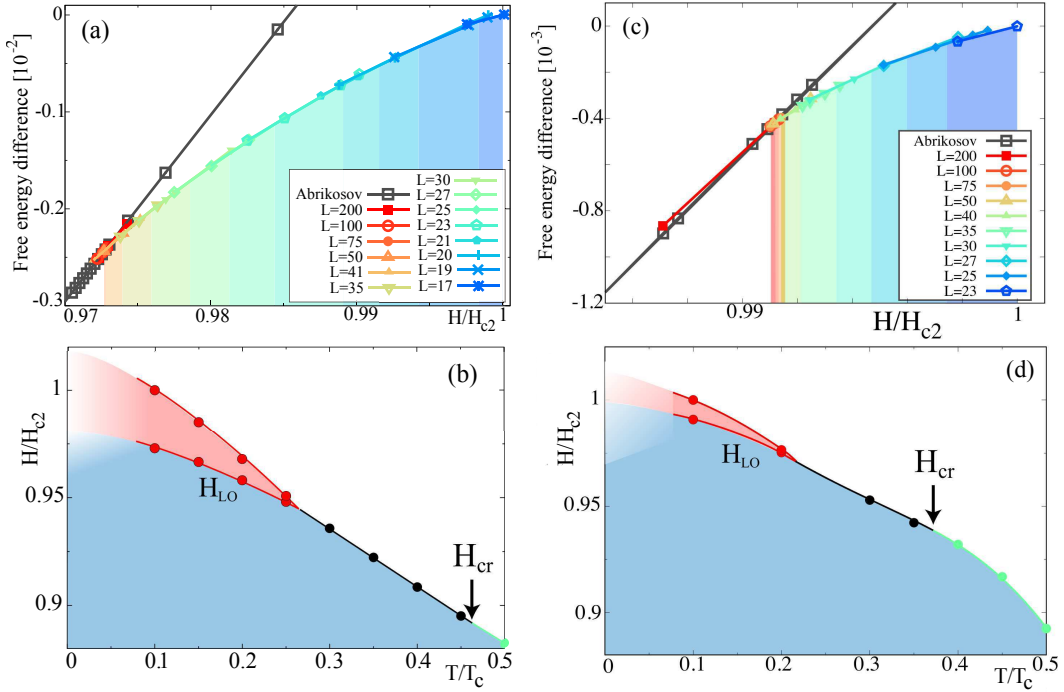} 
 \caption{ 
(Color online)  
(a) and (c) Free energies $F$ of the LO states with different wave numbers $L$
and Abrikosov state relative to the normal state as a function of $H$
for $\mu$= 5 and $\mu$= 2 respectively at $T = 0.1T_c$.
 (b) and (d) Phase diagrams for LO in $H$-$T$ plane for $\mu$= 5 and $\mu$= 2 respectively.
 $H$ is normalized by $H_{c2}$ at $T = 0.1T_c$.
 The upper red (lower blue) region  is the LO (Abrikosov) phase.
%The lower boundary $H_{\rm LO}$  is second order. 
$H_{\rm c2}$ is first order for $H>H_{\rm cr}$. 
Lines are guide for the eye. 
} 
\label{fig:phase-diagram1} 
\end{figure}

Figures \ref{fig:phase-diagram1}(a) for $\mu=5$ and 
\ref{fig:phase-diagram1}(c) for $\mu=2$ exhibit the
resulting successive changes of $F$ and $L$ at $T/T_{\rm c}=0.1$.
At around $H=H_{\rm c2}$ the LO with the
shortest wave length is stabilized which is $L\sim17$ for $\mu=5$
and  $L\sim23$ for $\mu=2$. Note that the length unit $R_0$ is roughly equal to
the coherence length $\xi_0$.
As $H$ decreases, $L$ becomes longer and longer. 
Eventually the free energy of the LO becomes comparable with that 
of the Abrikosov state 
where the LO modulation along the field direction is absent.
The envelop of the free energies of the LO approaches that
of the Abrikosov state, 
such that the two curves seem to merge tangentially, namely
 at the meeting point the tangents of the two curves coincide with 
 each other. 
%Of course our numerics  can not prove it because 
%$L$ is discretized.
%This is also true for the successive changes mentioned above.
%In the actual situation the wavelength $L$ should vary continuously
%and smoothly.  
While our calculations are done for discretized $L$, 
these results suggest:
second order like transition at $H_{\rm LO}$~\cite{ikeda} and the 
continuous
$L$ change as a function of $H$ in the LO state $H_{\rm LO}<H<H_{c2}$,
similar to results of previous analytic LO theory~\cite{nakanishi}.
$H_{\rm LO}$ is the transition field from the Abrikosov vortex state to 
the LO state.

We also notice here that as seen from Figs. \ref{fig:phase-diagram1}(a) and \ref{fig:phase-diagram1}(c)
the Abrikosov state shows the first order transition if the LO state is absent.
Then the LO states enhance the upper critical field $H_{\rm c2}$ substantially.
The superconducting state survives to higher fields by creating the LO states.
The enhancement is larger for $\mu=5$ than for $\mu=2$.

%%%%%%%%%%fig%%%%%%%%%%%%%% 
%%%%%%%%%%%%%%%%%%%%%%%%  

In Figs.~\ref{fig:phase-diagram1}(b) for $\mu=5$, and (d)
for $\mu=2$,
we show the resulting phase diagrams in the $H$-$T$ plane.
Those are obtained by repeating the LO calculations as a function of $H$ 
at different temperatures, 
$T/T_c$=0.1, 0.15, 0.2, and 0.25  for $\mu=5$, and $T/T_{\rm c}$=0.1 and 0.2 for $\mu=2$. 
We show the critical point ($T_{\rm cr}$, $H_{\rm cr}$) in Figs.~\ref{fig:phase-diagram1}(b) and (d). 
The transition at $H_{c2}$ to the normal state is first order at $H>H_{\rm cr}$.
%Although  Gunther and Grunberg\cite{gunther}
%show that the phase boundary between the LO and the normal state is of second order, it is quite
%difficult to prove or disprove it within our numerical method.
%\textcolor{red}{
It is seen that the strong paramagnetic case $\mu=5$ in Fig. \ref{fig:phase-diagram1}(b), 
the LO phase appears only near $H_{\rm c2}$, and 
$H_{\rm LO}$ increases on lowering $T$  
in this typical example of isotropic Fermi sphere. 
As for the $\mu=2$ case in Fig. \ref{fig:phase-diagram1}(d), the basic features of the phase diagram are essentially 
the same as $\mu=5$ except that the LO phase shrinks and becomes narrower.
The LO region in the $H$-$T$ plane is given by $H_{\rm LO}/H_{c2}=0.973$
for $\mu=5$ at $T/T_c$=0.1,
which depends on the $\mu$ value, namely,
$H_{\rm LO}/H_{c2}=0.991$ for $\mu=2$ at $T/T_c$=0.1.
%$H_{\rm LO}/H_{c2}=0.989$ at $T/T_c$=0.2. 
To obtain wider LO region, 
we have to consider the contribution of realistic Fermi surface shape
such as quasi 2D shape~\cite{ichiokaFFLO} for better nesting condition, 
or  multi-band effect~\cite{2bandFFLO,2bandFFLO2}.  

Those phase diagrams are different from those for the
Zeeman depairing without the orbital depairing~\cite{nakanishi} and also
for the neutral Fermi superfluids with spin imbalance~\cite{machida}.
In the former case $H_{c2}$ of the LO phase shifts to much higher fields while
much wider LO phase is obtained in the latter case.

We notice the 
canonical phase diagram~\cite{nakanishi}, consisting of the second order line at higher $T$,
which bifurcates into two second order lines at lower $T$
in the theory without considering first order transition.
In the present calculation, we show how the phase diagram changes in the presence of first order transition. 
The bifurcate point is known as the tricritical point, so-called Lifshitz point $H_{\rm Lifshitz}$~\cite{nakanishi,fujita}.
According to the canonical phase diagram~\cite{nakanishi,fujita}, it is expected that $H_{\rm cr}$ should coincide 
with the endpoint of the $H_{\rm LO}$ line,
namely, the LO phase starts from $H_{\rm cr}$.
It is known that $T_{\rm LO}/T_{\rm c2}=0.56$ in the limit of $\mu\rightarrow\infty$.
Thus we understand that  the $\mu=5$ case almost approaches the strong Pauli paramagnetic effect limit
because $T_{\rm cr}\sim0.48T_{\rm c2}$
and the $\mu=2$ case is intermediate because $T_{\rm cr}\sim0.38T_{\rm c2}$.

% could be a few percent at most within the present single band model. 
%\textcolor{blue}{
%Note that if there is other band in a system,
%the LO region can become wider because this band acts as a reservoir
%to supply excess electrons into the active band\cite{fujita}.}
%In the limit of the zero density of states case, $H_{\rm LO}/H_{c2}$
%tends to zero at $T=0$, that is, the LO phase occupies the whole mixed phase
%(see Fig. 4or5?? in ref. \citen{machida}).

%%%%%%%%%%fig%%%%%%%%%%%%%% 
\begin{figure}
\centering 
\includegraphics[width=6cm]{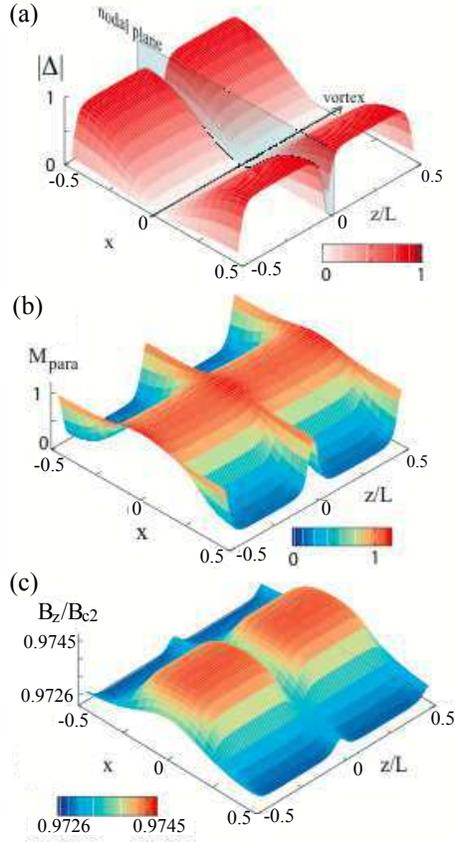} 
 \caption{ 
(Color online)  
Three dimensional spatial profiles of (a) the order parameter $|\Delta(x,z)|$ 
normalized by the maximum value $|\Delta(x=0.5c_x,z=0.25L)|$, 
(b) paramagnetic moment $M_{\rm para}(x,z)/M_0$,
and (c) induction field $B_z(x,z)$ normalized by  $B_{\rm c2}(T/T_c = 0.1) $. 
$T/T_c = 0.1$, $L=75$, $H$=0.973$H_{\rm c2}$ and $\mu=5$.
The nodal planes are situated at $z/L=-0.5,0, +0.5$ and the vortex center at $x=0$.
The profiles are displayed in one unit cell, 
$-0.5 \le x/c_x \le 0.5$ and $-0.5 \le z/L \le 0.5$. 
} 
\label{fig:profile} 
\end{figure} 
%%%%%%%%%%%%%%%%%%%%%%%%  

%%%%%%%%%%%%%%%%%%%%%%%% 
\begin{figure}
\centering
\includegraphics[width=7cm]{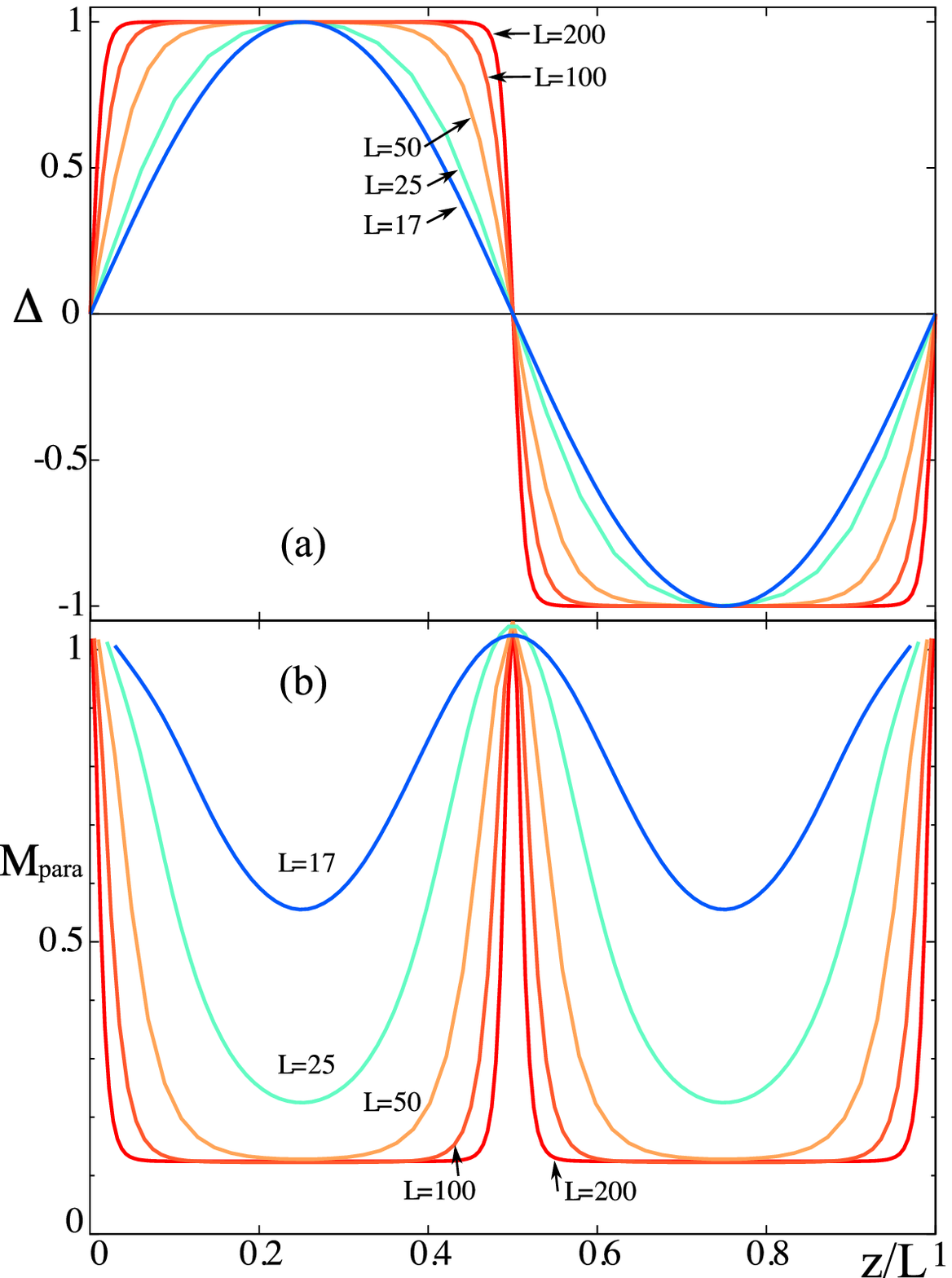}  
\caption{ 
(Color online)  
Cross-sectional views of (a) the order parameter $\Delta (x=\pm 0.5c_x,z)$ 
normalized by the maximum value $|\Delta(x=0.5c_x,z=0.25L)|$,
and (b) paramagnetic moment $M_{\rm para}(x=\pm 0.5c_x,z)/M_0$ 
for various wave numbers $L$ along field direction $z$ outside of vortex core region at $x=\pm 0.5c_x$ and $y=0$.
$T/T_c = 0.1$ and $\mu=5$. 
} 
\label{fig:pair-potential} 
\end{figure} 
%%%%%%%%%%%%%%%%%%%%%%%%  

\section{Spatial structure of the LO state}

We investigate the three dimensional spatial structures of various quantities in the LO states.
Figure~\ref{fig:profile}(a) displays the spatial profiles of the order parameter $\Delta(x,z)$
whose sign alternates along the $z$-direction. At $z$=0 the order parameter amplitude
vanishes where the paramagnetic moment $M_{\rm para}({\bf r})$ builds up
in addition to the vortex core at $x$=0 as shown in Fig. \ref{fig:profile}(b).
The LO nodal kink forms a sheet of the paramagnetic moments 
perpendicular to the field. 
Magnetic induction field $B_z$ is large along the vortex core at $x=0$ and suppressed at the
domain wall of the LO at $z=0$. 
These $B_z$ distributions indicate the confinement of $B_z$ at the vortex core is weak at the LO nodal line.
%On the other hand, 
%along the vortex lines, the enhanced $M_{\rm para}({\bf r})$ 
%at the vortex core is decreased at the intersection with LO kink plane
%(see Fig.2 in ref.~\onlinecite{ichiokaFFLO} and Fig.1 in ref.~\onlinecite{mizushima2}).
%There, as explained before~\cite{mizushima2}, the zero-energy peak states of quasiparticles are absent
%because of 2$\pi$ phase shift of the order parameter, 
%each coming from the kink and from the vortex. 

The paramagnetic moment  becomes strongly confined to the
kink position as $H$ approaches $H_{\rm LO}$ from the above.
This is seen also from Fig.  \ref{fig:pair-potential} more clearly.
These features of the three dimensional LO spatial structure
can be probed by SANS experiment or NMR experiment.

Figure \ref{fig:pair-potential} shows
the cross-sectional views of the normalized wave forms of the order parameter $\Delta(x=\pm 0.5c_x,z)$ (a) 
and paramagnetic moment $M_{\rm para}(x=\pm 0.5c_x,z)$ (b) in LO states 
along the field direction outside of vortex core region where
$x=\pm 0.5c_x$ are midpoints between nearest neighbor vortices. 
It is seen that a simple sinusoidal modulation wave form for $L=17$ stabilized near $H_{\rm c2}$ continuously
deforms into an anti-phase kink form, or solitonic wave form as 
$H$ approaches $H_{\rm LO}$ line, at which $L$ diverges~\cite{nakanishi,fujita}.
In other words near the $H_{\rm LO}$ boundary, 
the sign change or $\pi$ phase shift of the order parameter occurs sharply. 
For the longer $L$ near $H_{\rm LO}$ due to the excess normal electrons $M_{\rm para}({\bf r})$ 
is confined in a narrow spatial region along the kink position
as clearly seen from Fig.~\ref{fig:pair-potential}(b). 
For shorter $L$ approaching $H_{c2}$, $M_{\rm para}({\bf r})$  is changed to sinusoidal wave form. 
These changes of the LO structure reflect  to the behaviors of FLL form factors and NMR spectra, as discussed later.

\section{Field evolutions of thermodynamic quantities}

Thermodynamic quantities, such as magnetization curve, $M_{\rm para} (H)$, and the Sommerfeld coefficient $\gamma(H)$ 
under the Pauli paramagnetic effect in the Abrikosov state are evaluated in previous studies~\cite{amano,amano2,machida2,214nakai}.
Here we continue those into the LO state, which takes over the Abrikosov phase in higher fields.

%%%%%%%%%%%%%%%%%%%%%%%% 
%\begin{figure}[tb]
\begin{figure}  
\centering
 \includegraphics[width=14cm]{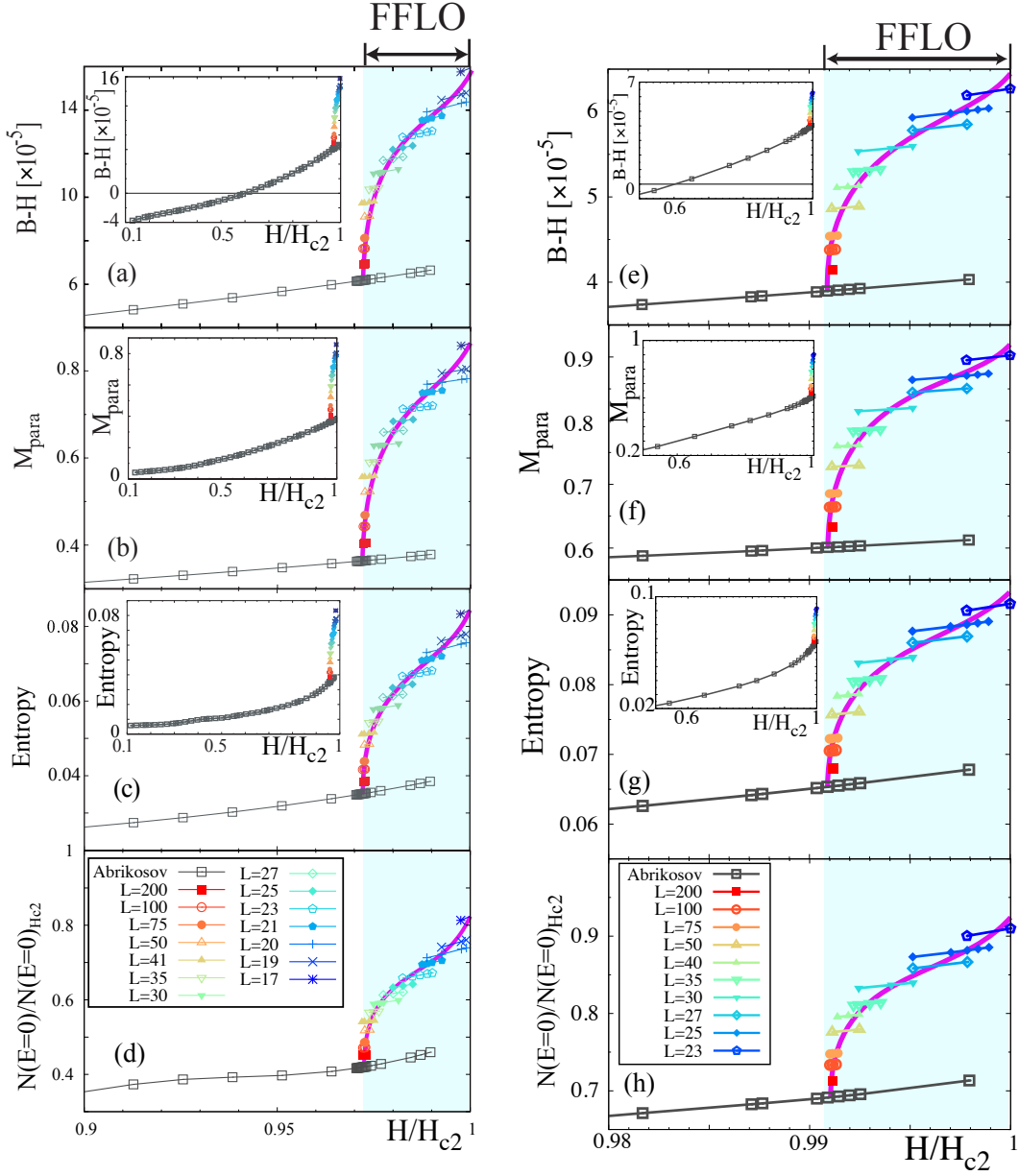}
\caption{ 
(Color online) 
 (a) and (e) Magnetic field $H$ dependence of magnetization $M=B-H$, 
 (b) and (f) paramagnetic susceptibility 
$\chi_{\rm spin}=M_{para}/M_0$ , 
 (c) and (g) entropy $S_{\rm s}(T)/S_n(T_{\rm c})$,
and (d) and (h) zero energy DOS $\gamma=N(E=0)/N_0$. 
The left panels (a)-(d) are for $\mu$=5, 
and right panels (e)-(h) for $\mu$=2. 
$T=0.1T_{\rm c}$.  
Insets in upper three panels show the overall features in wide range of $H$ from low fields. 
Data points are plotted for $L$ near the free energy minimum, 
with color presented in the lowest panels. 
Continuous curves are drawn for guide of the eye.
} 
\label{fig:thermo1} 
\end{figure} 
%%%%%%%%%%%%%%%%%%%%%%%%  

\subsection{Magnetization}

Figures \ref{fig:thermo1}(a) and \ref{fig:thermo1}(e) show magnetization curve $M=\bar{B}-H$ 
at $T=0.1T_{\rm c}$ for $\mu=5$ and $\mu=2$ respectively. 
The magnetization $M$ includes paramagnetic and diamagnetic contributions. 
As is seen in the insets, $M <0$ at low $H$ as the diamagnetic contribution is dominant. 
If the Pauli paramagnetic effect is absent, $M \rightarrow 0$ when $H \rightarrow H_{\rm c2}$. 
However, in the presence of the Pauli paramagnetic effect, $M$ becomes positive at high fields 
since the paramagnetic component $M_{\rm para}$ becomes dominant. 
Due to the larger paramagnetic contribution, $M$ is larger for $\mu=5$, 
compared with that for $\mu=2$. 
In the Abrikosov state below $H < H_{\rm LO}$, $M$ monotonically increases with a slow slope. 
When the Abrikosov state is changed to the LO state at $H>H_{\rm LO}$, 
we see a  rapid increase of $M$. 
In Fig. \ref{fig:thermo1}, we plot data points for some $L$ near free energy minimum. 
The continuous curves are drawn for guide of the eye. 
On the curves, the diverging slope at $H_{\rm LO}$ is gradually changed to a slower slope both for $\mu=5$ and 2. 
In the narrow field region $H_{\rm LO}< H < H_{\rm c2}$, $M$ increases toward the normal state value $M_0$ at $H>H_{\rm c2}$. 
The increase is larger for larger $\mu$. 
We see a small jump of $M$ at $H_{\rm c2}$. 
Although we expect a large jump of $M$ at $H_{\rm c2}$ in the Abrikosov state if the LO state is absent, 
the jump is smeared by the increase of $M$ due to the presence of the LO state at $H_{\rm LO}< H < H_{\rm c2}$.

 \subsection{Paramagnetic susceptibility}

The $H$ dependence of the normalized paramagnetic susceptibility $\chi_{\rm spin}=M_{\rm para}/M_0$ is presented in Figs. \ref{fig:thermo1}(b) and \ref{fig:thermo1}(f) for $\mu=5$ and $\mu=2$ respectively.  
The extrapolation of lines for $\chi_{\rm spin}$ in the Abrikosov state toward higher $H$ until $\chi_{\rm spin}=1$ suggests the orbital limit of $H_{\rm c2}$. 
The higher $H_{\rm c2}$ of the orbital limit is suppressed by the Pauli paramagnetic effect, 
and $\chi_{\rm spin}$ shows jump at the first order $H_{\rm c2}$ transition, 
as is shown in Fig.~\ref{fig:thermo1}(b) and Fig.~\ref{fig:thermo1}(f).  
The jump is larger for larger $\mu$. 
Since the dominant contribution of $M$ comes from 
the paramagnetic part $M_{\rm para}$ at high fields, 
$M_{\rm para}$ in Figs. \ref{fig:thermo1}(b) and \ref{fig:thermo1}(f) shows similar behavior to $M$ in 
Figs. \ref{fig:thermo1}(a) and \ref{fig:thermo1}(e) in the LO state. 
$\chi_{\rm spin}$ also shows a large increase in the LO state at $H_{\rm LO}< H < H_{\rm c2}$, and small jump to $\chi_{\rm spin}=1$ at $H_{\rm c2}$. 
In the LO state, $\chi_{\rm spin}$ changes from 0.37 at $H_{\rm LO}$ to 0.86 at $H_{\rm c2}$ for $\mu=5$ in  Fig.~\ref{fig:thermo1}(b)
and from 0.6 to 0.9 for $\mu=2$ in  Fig.~\ref{fig:thermo1}(f).  

 \subsection{Entropy}

The $H$-dependence of entropy $S_{\rm s}(T)/S_{\rm n}(T_{\rm c})$ is presented in Figs. \ref{fig:thermo1}(c) and \ref{fig:thermo1}(g) for $\mu=5$ and $\mu=2$ respectively. 
These behaviors show similar $H$-dependence as in $\chi_{\rm spin}$ in Fig.~\ref{fig:thermo1}(b) and Fig.~\ref{fig:thermo1}(f). 
The entropy also shows rapid increase in the LO state at 
$H_{\rm LO} < H < H_{\rm c2}$, and small jump to the normal state value 0.1 at $H_{\rm c2}$. 
In the LO state, $S_{\rm s}(T)/S_{\rm n}(T_{\rm c})$ changes from 0.035 to 0.084 for $\mu=5$ in 
Fig.~\ref{fig:thermo1}(c), and from 0.065 to 0.092 for $\mu=2$ in Fig.~\ref{fig:thermo1}(g). 
Quantitatively,  
$S_{\rm s}(T)/S_{\rm n}(T_{\rm c})$ is smaller by a factor of about 0.1 $(=T/T_{\rm c})$. 
Compared with $\chi_{\rm spin}$, $S_{\rm s}$ shows small enhancement near $H_{\rm LO}$ in the Abrikosov state
 as seen in the insets.  

 \subsection{Zero-energy DOS}
 
Figures \ref{fig:thermo1}(d) and \ref{fig:thermo1}(h) show the $H$-dependence of the zero-energy 
DOS $N(E=0)$, which also shows similar behavior to that of $\chi_{\rm spin}$ and $S_{\rm s}$ in the above panels in Fig. \ref{fig:thermo1}. 
The thermodynamic quantity also strongly increases with almost diverging slopes at $H_{\rm LO}$. 
In the LO state, $N(E=0)/N_0$ changes from 0.42 to 0.82 for $\mu=5$ in Fig.~\ref{fig:thermo1}(d), and from 0.69 to 0.91 for $\mu=2$ in Fig.~\ref{fig:thermo1}(h). 
The specific heat $C$ is obtained by the derivative of $S_{\rm s}(T)$ as 
%%%
\begin{eqnarray}
C=T \frac{\partial S_{\rm s}} {\partial T}. 
\label{eq:C}
\end{eqnarray}
We note here that in the low temperature limit
$C$ is evaluated as 
\begin{eqnarray}
\left({C\over T}\right)_{T\rightarrow 0}
=\left(\frac{\partial S_{\rm s}} {\partial T} \right)_{T\rightarrow 0} 
=\frac{S_{\rm s}(T)- S_{\rm s}(0)}{T-0}=\frac{S_{\rm s}(T)}{T}, \ 
\label{eq:C=S}
\end{eqnarray}
that is, the Sommerfeld coefficient $\gamma(H)=C/T$ is directly related to the entropy,
\begin{eqnarray}
\gamma(H)={S_{\rm s}(H)\over T}.
\label{eq:gamma}
\end{eqnarray}
at the low $T$ limit. 
We roughly confirm this relation  
from the numerical results of $S_{\rm s}(H)$ and $\gamma(H)$ at $T=0.1T_{\rm c}$ in Fig.~\ref{fig:thermo1}. 
The small deviations between them come from the effects of finite $T$. 
We also approximately confirm the relation $\chi_{\rm spin}(H) \sim \gamma(H)$ in Fig.~\ref{fig:thermo1}. 
This relation is confirmed also in the LO state in addition to the Abrikosov state, which was proved for the latter state in previous studies~\cite{amano,machida2,ichioka}.
Although the calculation of $\chi_{\rm spin}(H)$ is performed by Matsubara frequency $\omega_n$, in the formulation of real energy $E$, 
$\chi_{\rm spin}(H)$ comes from the average of the DOS in the energy range 
$|E| < \mu H$ at low $T$. 
Thus, we have the relation $\chi_{\rm spin}(H) = \gamma(H)$ in the limit of 
weak Pauli paramagnetic effect, $\mu \rightarrow 0$, and low $T$. 
When $\mu$ is large, the deviation may appear between $\chi_{\rm spin}(H)$ and $\gamma(H)$.

As is seen above, we confirmed that thermodynamic quantities of magnetization, 
paramagnetic susceptibility, entropy, and low temperature specific heat exhibit basically similar behaviors as a 
function of $H$. Namely
as $H$ increases, the almost linear and monotonic increase suddenly shows a sharp
rise at $H=H_{\rm LO}$ exhibiting a kink feature, but the thermodynamic quantities are continuous.
Thus it is of second order transition.
This feature nicely corresponds to that in the analytic solutions~\cite{nakanishi,fujita}, 
where at the tricritical Lifshitz point $L$ diverges from the above.  

Although it is difficult to check whether it is second or first order transition,
it is believed to be second order, judging from the analytic solutions~\cite{nakanishi,fujita}.
However, it often happens that the actual experiments show the first order transition
because of other degrees of freedom such as phonons or lattice deformation involved.
As for the phase transition at $H_{\rm c2}$,  the rise terminates at $H=H_{\rm c2}$ 
abruptly via first order like jump.

Comparing the two cases for $\mu=5$ (left column) and $\mu=2$ (right column)
in Fig.~\ref{fig:thermo1}, it is seen that the former has a wider LO region than the
latter. Otherwise, the two cases are quite similar, meaning that the qualitative features of the LO phase are
independent of the $\mu$ parameter and thus universal. As $\mu$ decreases, the LO phase fades out from the $H$-$T$ plane.
Note that the critical $\mu$ is known to be $\mu_{\rm cr}=0.5$.
Those thermodynamic quantities are expected to be measured by a variety of experiments, such
as the specific heat at low $T$ directly probes $N(0)$ and entropy.
The paramagnetic moment is measured directly by magnetization experiment,
which was conducted in CeCoIn$_5$, giving similar overall characteristics~\cite{tayama} 
shown in Figs.~\ref{fig:thermo1}(a) and~\ref{fig:thermo1}(e)
or by SANS experiment through diffraction of the spatial variation of magnetization profile~\cite{Tm}.

\section{FLL form factors}

\subsection{Period $L(H)$ in the LO state}

We first show the field evolution of the period $L$ or the wave number  $q=2 \pi/L$ 
of the LO state before discussing the FLL form factors.
As shown in Fig.~\ref{fig:SANS1}(a) for $\mu=5$ and Fig.~\ref{fig:SANS1}(e) for $\mu=2$, 
the wave number $q$ of the stable LO state continuously varies with $H$.
Starting with $q=0$ at $H=H_{\rm LO}$ where the LO period is infinity, $q$ rises sharply
whose tangent is almost diverging.  Thus $L$ becomes finite quickly.
The anti-phase solitonic-wave form changes 
into a sinusoidal one upon increasing $H$ (see also Fig.~\ref{fig:pair-potential}(a)). 
This behavior is similar to that seen 
in the exact solution (see Fig. 9 in Ref.~\onlinecite{nakanishi}), 
implying that the LO physics along the parallel direction
exemplified here is common and universal, which was also pointed out in Ref.~\onlinecite{tachiki}.
Comparing with the two cases $\mu=5$ in Fig.~\ref{fig:SANS1}(a) and $\mu=2$ in Fig.~\ref{fig:SANS1}(e)
the $q(H)$ variation is somewhat rounded in $\mu=2$.
%\textcolor{blue}{
%Needless to say,
%\textcolor{red}{these behavior are completely different from those in the vertical plane, 
%there the vortex lattice is formed.\cite{ichiokaFFLO}}}

\subsection{Fundamental form factor $F_{100}$}

\begin{figure}
\centering
 \includegraphics[width=14cm]{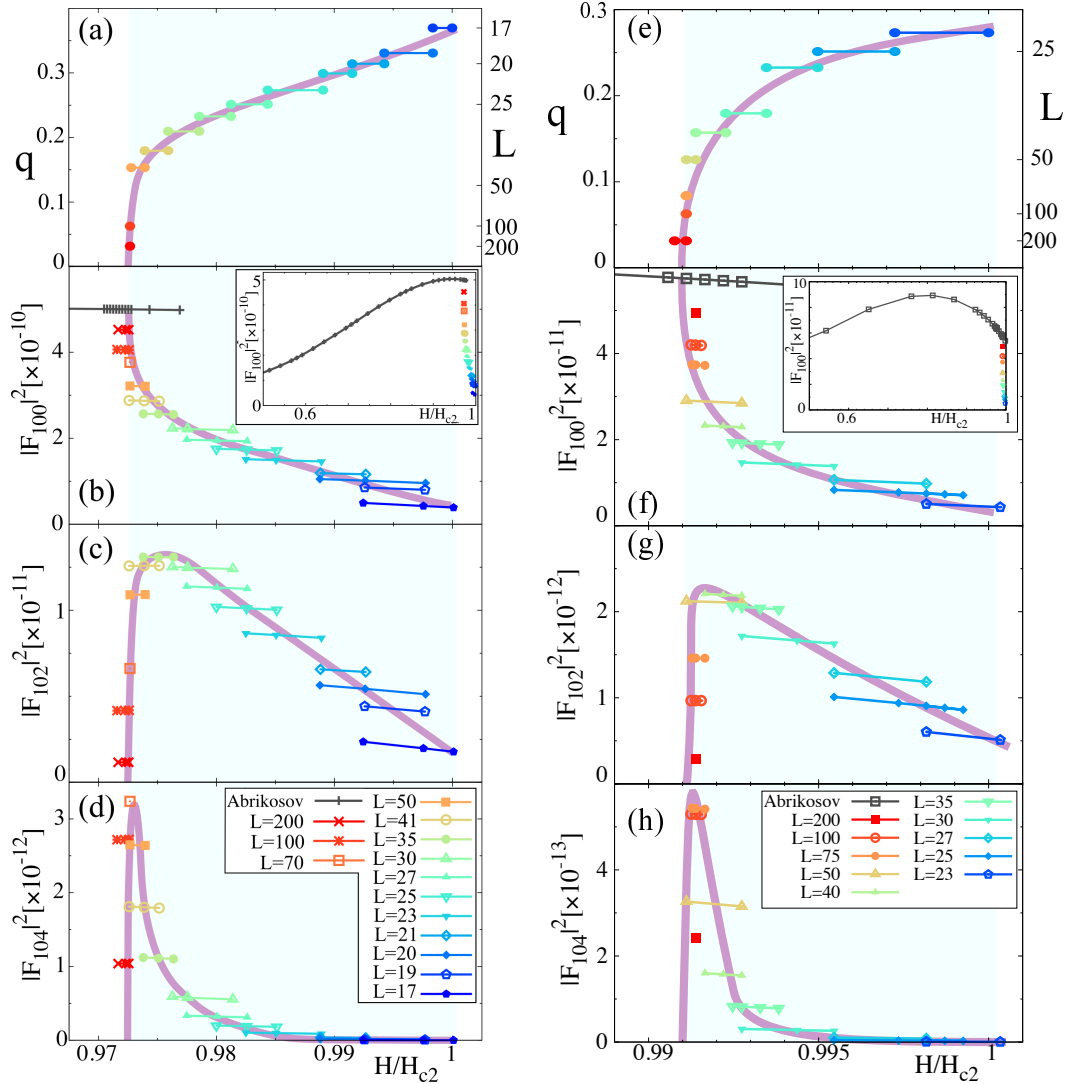}
\caption{ 
(Color online)  
Field evolutions of various quantities at $T/T_c=0.1$
for $\mu=5$ (left column) and $\mu=2$ (right column). 
 (a) and (e) LO wave number $q=2 \pi/L$.
 (b) and (f) Form factor $|F_{100}|^2$. Inset shows the overall variation.
 (c) and (g) Form factor $|F_{102}|^2$.
 (d) and (h) Form factor $|F_{104}|^2$.
Data points are plotted for $L$ near the free energy minimum, 
with color presented in the lowest panels. 
Continuous curves are drawn for guide of the eye.
} 
\label{fig:SANS1} 
\end{figure}

The FLL form factor is an important quantity that  can be directly measured by 
SANS experiment. 
The form factors $F_{hkl}$ with $h,k$, and $l$ being integers are Fourier components 
of internal field ${\bf B}({\bf r})$ 
in our calculation~\cite{ichiokaFFLO}. 
The fundamental Bragg spots $F_{100}$ for the vortex lattice 
is shown in Fig.~\ref{fig:SANS1}(b) for $\mu=5$ and Fig.~\ref{fig:SANS1}(f) for  $\mu=2$
 as a function of $H$.
The intensity $|F_{100}|^2$ increases in the Abrikosov state as seen from the insets
in Figs.~\ref{fig:SANS1}(b)  and~\ref{fig:SANS1}(f).  
This is because $M_{\rm para}({\bf r})$ accumulates at the vortex core
to increase $B({\bf r})$ locally.
This feature is already shown theoretically\cite{ichioka} 
and observed in various paramagnetically enhanced superconductors,
such as in TmNi$_2$B$_2$C\cite{Tm} and  CeCoIn$_5$\cite{morten}.
While the increase of $|F_{100}|^2$ as a function of $H$ is greater for $\mu=5$ of strong Pauli paramagnetic 
effect case, for the intermediate case $\mu =2$, $|F_{100}|^2$ shows decrease at higher fields after the increase at lower fields. 
 
As shown in the main panels of Figs.~\ref{fig:SANS1}(b), and \ref{fig:SANS1}(f), 
the intensity of $|F_{100}|^2$ suddenly decreases upon entering the
LO phase and keeps dropping quickly almost exponentially (Notice the $T=50$ mK data in Fig.1 of Ref.~\onlinecite{white}).
This is because $B_{z}({\bf r})$ 
is not enhanced at the vortex core on the LO nodal plane
%transfer from vortex cores to LO nodal planes.}
%\textcolor{red}{$\to$ 
%This is because in the nodal plane emerged $M_{\rm para}$ 
%make its structure from vortex lattice smooth
%
%In the LO nodal plane $M_{\rm para}$ emerge everywehere.
%This makes structure of $M_{\rm para}$ from vortices smooth
%and leads to .
as seen from Fig.~\ref{fig:profile}(c).
This contribution decreases $|F_{100}|^2$ which is the average along
the $z$-axis.
Comparing with the two cases $\mu=5$ in Fig. ~\ref{fig:SANS1}(b) and $\mu=2$ in Fig.~\ref{fig:SANS1}(f),  
the $|F_{100}|^2$ variation in the LO state is somewhat rounded in  $\mu=2$, similarly to the $q(H)$ behavior 
in Figs.~\ref{fig:SANS1}(a) and ~\ref{fig:SANS1}(e). 
This indicates that the decrease of $|F_{100}|^2$ in the LO state is related to $q(H) =2 \pi /L$, i.e., 
volume weight of the LO nodal sheet in the superconductor. 
The other Bragg spots $F_{hk0}$ ($h,k$ integers) are
associated with the vortex lattice, which characterize the detailed magnetic field distribution
in the mixed state of a superconductor.

%%%%%%%%%%%%%%%%%%%%%%%% 
%%%%%%%%%%%%%%%%%%%%%%%%  

\subsection{Form factors $F_{102}$ and $F_{104}$ associated with LO state}

The observation of extra spots $F_{10n}$ ($n=$2,4, ...) 
is crucial to prove the existence of the LO phase. 
In Figs.~\ref{fig:SANS1}(c) and~\ref{fig:SANS1}(g)
we show $|F_{102}|^2$ that is the superspot associated with 
the LO modulation along the field direction. 
$|F_{102}|^2$ rises sharply at $H=H_{\rm LO}$.
After taking a maximum in the middle of the LO phase, 
it slowly decreases toward $H_{\rm c2}$.
Note that $|F_{102}|^2$ behaves similarly for both $\mu=5$ and $\mu=2$ cases.
Thus the results may not be sensitive to the $\mu$ value and generic.
The best chance to observe $|F_{102}|^2$ superspot is in the middle field region 
inside the LO phase.
The relative intensity $|F_{102}|^2/|F_{100}|^2= 1/10 \sim 1/20$ in both $\mu=5$ and $\mu=2$.
It is possible to detect the $F_{102}$ spot
because $|F_{100}|^2$ is enhanced by the Pauli paramagnetic effect even near $H_{\rm c2}$.

The higher order spot $|F_{104}|^2$ is also shown in Figs.~\ref{fig:SANS1}(d) 
and~\ref{fig:SANS1}(h). 
It takes a maximum just near $H_{\rm LO}$. 
Since the magnitude of $|F_{104}|^2$ is further reduced compared with $|F_{102}|^2$
and is one order of magnitude smaller than $|F_{102}|^2$,
it might be difficult to detect $|F_{104}|^2$.
The $\mu$ parameter dependence of those form factors is qualitatively the same,
only differing quantitatively.

From Figs.~\ref{fig:SANS1}(c) and~\ref{fig:SANS1}(d) for $\mu=5$, with increasing $H$ from $H_{\rm LO}$, the ratio $|F_{102}/F_{104}|^2$ is evaluated as $1.3 \times 10^{-12} / 1.0 \times 10^{-12} =1.3 $ for $L=200$,  $6.8 \times 10^{-12} / 3.2 \times 10^{-12} =2.1$ for $L=70$, and $1.3 \times 10^{-11} / 1.1 \times 10^{-12} =12 $ for $L=35$. From Figs.~\ref{fig:SANS1}(g) and~\ref{fig:SANS1}(h) for $\mu=2$,  the ratio $|F_{102}/F_{104}|^2$ is $2.8 \times 10^{-13} / 2.4 \times 10^{-13} =1.2 $ for $L=200$,  $1.5 \times 10^{-12} / 5.4 \times 10^{-13} =2.8 $ for $L=75$, 
and $2.2 \times 10^{-12} / 1.6 \times 10^{-13} =14 $ for $L=40$.  
Both for $\mu=5$ and $\mu=2$, the ratio $|F_{102}/F_{104}|^2$ rapidly increases from 1 at $H > H_{\rm LO}$. 
At higher $H$, as $F_{104}$ becomes negligible, $B_z$ distribution becomes a sinusoidal wave of $F_{102}$ along the $z$ direction.

\section{NMR spectrum}

In this section we examine 
the NMR spectrum which is also crucial to identify the LO state.
Choosing probed nuclei that have different hyperfine coupling constants, 
we can measure the field distributions inside a supercondutor~\cite{ichiokaFFLO}.  
When the hyperfine coupling is strong enough, 
the paramagnetic distribution $M_{\rm para}({\bf r})$ is probed by NMR experiment.
In the weak hyperfine coupling case the magnetic induction $B({\bf r})$ 
in the whole system is detected by NMR.
In the mixed state of ordinary superconductors it 
yields the so-called Redfield pattern.
Here we analyze the field evolution of the NMR spectra both for strong and weak
hyperfine coupling cases. For the former we evaluate the distribution $P(M)$
by using the stable LO state determined at each field. And
for the latter  the distribution $P(B)$ is calculated.

%%%%%%%%%%%%%%%%%%%%%%%%%%%%%%%%%%%%%%%%%%%% 
\begin{figure}
\centering
 \includegraphics[width=7cm]{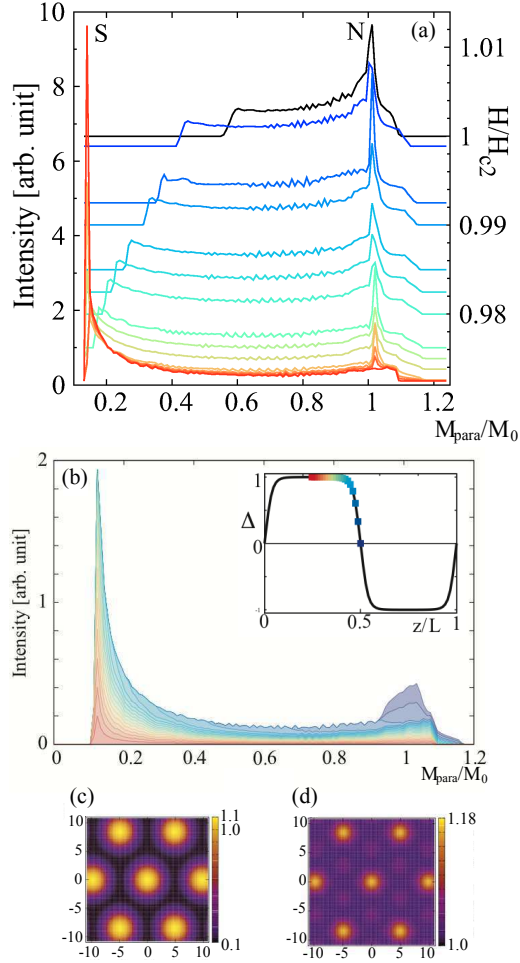}
\caption{ 
(Color online)
NMR spectra $P(M)$ in the LO state: 
(a) Applied field $H$ evolution of internal field distribution $P(M)$. 
$\mu=5$ and $T/T_c=0.1$.
Horizontal baselines for each spectrum are shifted by $H/H_{\rm c2}$, 
which is indicated on the right axis. 
(b) The $z$-resolved the paramagnetic moment $M_{\rm para}$ distribution $P(M)$.  
Inset  shows the order parameter profile as a function of $z$
where color codes correspond to those in the main figure.
(c) and (d) Density plots of paramagnetic moment $M_{\rm para}$ at the antinodal plane $z=0.25L$ and  at the nodal plane $z=0.5L$, 
respectively.
$T/T_c = 0.1$, $L=75$, $H$=0.973$H_{\rm c2}$ and $\mu=5$ for (b), (c), and (d).  
} 
\label{fig:PM} 
\end{figure} 
%%%%%%%%%%%%%%%%%%%%%%%%%%%%%%%%%%%%%%%%%%%%%%%%%  

\subsection{Paramagnetic distribution spectrum $P(M)$}

We start with the strong hyperfine coupling constant case,
which effectively probes the paramagnetic distribution $M_{\rm para}({\bf r})$
in the system. The distribution $P(M)$ is given by

\begin{eqnarray} 
P(M)=\left\langle \delta(M-M_{\rm para}({\bf r})) \right\rangle_{\bf r}, 
\end{eqnarray} 

\noindent
i.e. the volume counting for each $M$. 
Figure~\ref{fig:PM}(a)
% and Fig.~\ref{fig:PB}(a) 
shows the spectral evolutions of the distribution $P(M)$. 
 %and  $P(B)$ respectively.
Since in the Abrikosov state the paramagnetic moment is confined exclusively 
at the vortex cores, the single peak appears at the saddle point (S) position in the NMR spectrum.
In the LO phase, $M_{\rm para}({\bf r})$,  
which comes from excess electrons at the nodal sheets, accumulates near the normal state (N) position $M/M_0=1$. 
The peak near N-position becomes dominant toward $H_{\rm c2}$,  
because the increasing excess unpaired quasi-particles 
appear at the LO nodal sheets as described above. 
It is noticed that just near $H=H_{\rm LO}$ two peaks with nearly equal height appear in 
the NMR spectrum in $P(M)$, and the noticeable spectral weight is seen 
at the higher $M_{\rm para}$ region.  In addition to those characteristics,
the spectral weight extends to higher values beyond $M_0$ near $N$ positions.
This comes from the increase of the domain wall contributions in the LO state as discussed below.
Those features are important to characterize the spectra
nearthe N position in the LO state as shortly see in the last section.
%The double peak structure  is observed in In(2) site of the NMR spectrum 
%by Kumagai {\it et al.}~\cite{kumagai2} for CeCoIn$_5$.

The appearance of the double peaks at the S- and near the N-positions gives unambiguous
evidence of the LO state.
It may be possible to extract the wave length $L$ in the LO state
by carefully examining the spectral evolution data because
the spectral weights at S and N evolve continuously and gradually.
In order to understand the physical meanings of those spectra $P(M)$ in Fig.~\ref{fig:PM}(a) more deeply,
we examine the $z$-resolved $P(M)$ shown in Fig.~\ref{fig:PM}(b). 
%and $P(B)$ in Fig.~\ref{fig:PB}(a).
%Figure~\ref{fig:PM}(b)  
%and Fig.~\ref{fig:PB}(b) 
%shows the $z$-resolved $P(M)$ 
%and $P(B)$ respectively 
There
the bulk superconducting contribution at the S-point comes exclusively from the
maximum position near $z=0.25L$ of the order parameter amplitude. The 
normal contribution near the N-point arises from the nodal plane at the middle $z=0.5L$.
 The spectral distribution continuously evolves, depending on the order parameter 
 spatial variation. The prominent double horn structure is a hallmark
 of the LO state and the spectral weights at the S and N-points change, reflecting
 the field evolution of the LO state. Thus we can extract the information on the detailed LO spatial structure
 by carefully measuring the NMR spectrum.

As shown in Figs.~\ref{fig:PM}(c) and ~\ref{fig:PM}(d) 
%and Fig.~\ref{fig:PB}(c)((d)),
the cross-sectional views of the $M$ 
%and $B$ 
profile at the antinodal plane and nodal plane respectively are displayed.
Comparing those two cross-sectional views, it is seen that the
vortex core contrast relative to the background is far clear at the antinodal
plane than that at the nodal plane. This is because the latter contrast is blurred 
by normal quasi-particles accumulated at the nodal plane.
Note that the color range is 1.0$ < M/M_0< $1.18 in Fig.~\ref{fig:PM}(d) while 0.1$< M/M_0 < $1.1 in Fig.~\ref{fig:PM}(c). 
We point out here that according to the recent STM measurement~\cite{hanaguri} on
FeSe, which is a candidate material for the LO, under the perpendicular field to
the surface the vortex images become suddenly invisible and bluer when entering 
the possible LO phase. This phenomenon can be understood in the following:
At the surface where STM probes the electronic structure the nodal sheets are likely pinned
there because of energetic consideration, thus as shown in Figs.~\ref{fig:PM}(b) and~\ref{fig:PM}(c)
 the contrast at the nodal sheet 
is by far lower than that at the antinodal plane. Since the paramagnetic moment is
proportional to the DOS $N(E=0)$, we anticipate that the same is happening for STM
zero bias images.

%%%%%%%%%%%%%%%%%%%%%%%%%%%%%%%%%%%%%%%%%%%%% 
\begin{figure}[tb]  
\centering
 \includegraphics[width=7cm]{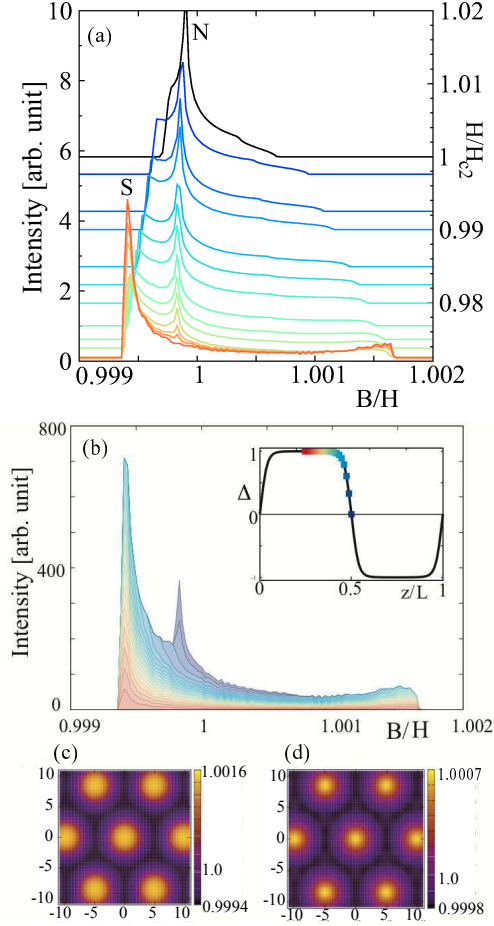}
\caption{ 
(Color online)
NMR spectra $P(B)$ in the LO state: 
(a) Applied field $H$ evolution of internal field distribution $P(B)$.  
$\mu=5$ and $T/T_c=0.1$.
Horizontal baselines for each spectrum are shifted by $H/H_{\rm c2}$, 
which is indicated on the right axis. 
(b) The $z$-resolved internal field $B$ distribution $P(B)$.
Inset  shows the order parameter profile as a function of $z$
where color codes correspond to those in the main figure.
(c) and (d) Density plots of internal field $B$ at the antinodal plane $z=0.25L$ and  at the nodal plane  $z=0.5L$, respectively.  
$T/T_c = 0.1$, $L=75$, $H$=0.973$H_{\rm c2}$ and $\mu=5$ for (b), (c), and (d). 
} 
\label{fig:PB} 
\end{figure} 
%%%%%%%%%%%%%%%%%%%%%%%%%%%%%%%%%%%%%%%%%%%%%%%%%%%%%  

\subsection{Magnetic induction distribution spectrum $P(B)$}

Next we study the weak hyperfine coupling constant case, which probes effectively 
the magnetic induction distribution $P(B)$ in the whole system. The distribution $P(B)$ is given by

\begin{eqnarray} 
P(B)=\left\langle \delta(B-B({\bf r}))  \right\rangle_{\bf r}.
\end{eqnarray} 

\noindent
It is also important to observe the characteristic change of $P(B)$
as shown in Fig.~\ref{fig:PB}(a).
Note that $P(B)$ is probed for example, at In(1) in CeCoIn$_5$~\cite{kumagai2}.
The double peak structure can be seen from
Fig.~\ref{fig:PB}(a) in the LO phase at $H>H_{\rm LO}$, 
where the $N$ peak appears near $B\sim H$ in the spectrum.
Viewing the whole spectral shape in Fig.~\ref{fig:PB}(a), the N-position is situated near the S-position in $P(B)$, 
compared with $P(M)$ in Fig.~\ref{fig:PM}(a). 
In the lower field of the Abrikosov state, the usual Redfield pattern 
is reproduced as seen from Fig.~\ref{fig:PB}(a).
Thus the double peak structure at the $N$ and $S$ positions in $P(B)$ is a hallmark of the LO state.
As $H$ increases the relative spectral weight changes and eventually the spectral weight at $N$ dominates the
whole spectrum toward $H_{c2}$, which is shown in Fig.~\ref{fig:PB}(a).
Those eminent features of the NMR spectra in $P(B)$ can be useful and indispensable 
spectroscopic methods for identifying the LO state.
Furthermore it may be possible to extract the details of the LO state, such as the LO periodicity,
by carefully examining those spectra.

As is shown in Fig.~\ref{fig:PB}(b) the double peak structure is analyzed by 
decomposing the spectral weight into the $z$-resolved $P(B)$. 
The peak of the S-position comes from the contributions of the antinodal parts around  $z/L=0.25$ while that of the N-position comes from the
LO nodal sheet at $z/L=0.5$ as seen in the inset of Fig.~\ref{fig:PB}(b).

The cross-sectional views at the antinodal and nodal positions are displayed in Figs.~\ref{fig:PB}(c) and~\ref{fig:PB}(d)
respectively.
It is seen by comparing the scales that the contrast of the spectral weight at the antinodal plane in 
Fig.~\ref{fig:PB}(c) is far visible than at
the nodal plane in Fig.~\ref{fig:PB}(d). This is the same as in the $P(M)$ case mentioned above.

%%%%%%%%%%%%%%%%%%%%%

\section{Discussions}

Having calculated various physical properties of the stable LO states in detail,
we now examine the possible experiments to identify the LO phase 
in several candidate materials, SrRuO$_4$, CeCoIn$_5$, CeCu$_2$Si$_2$, and the organic
superconductors (BEDT-TTF)$_2$X in light of the present theory.

\subsection{Sr$_2$RuO$_4$}

Sr$_2$RuO$_4$ was a prime candidate of the chiral $p$-wave superconductor. Much attention was focused on this
symmetry. However, the recent trends, including (1) First order transition at $H^{ab}_{\rm c2}\parallel ab$
found by magnetocaloric effect~\cite{kajikawa},  specific heat~\cite{yonezawa} and magnetization experiments~\cite{kittakaM},
(2) The intrinsic anisotropy 60 observed by SANS~\cite{morten214,kuhn} as a vortex lattice deformation
indicates that $H_{\rm c2}$ anisotropy $H^{ab}_{\rm c2}/H^{c}_{\rm c2}=20$
is a suppressed value by the Pauli paramagnetic effect, (3) Absence of the split transition under uniaxial stresses~\cite{hicks}, 
expected for chiral $p$-wave pairing belonging
to a two-dimensional irreducible representation, and finally (4) The renewed Knight shift experiment~\cite{stuart} detects 
a decrease of the spin susceptibility below $T_c$ for $H\parallel ab$.  This demonstrates that 
the original results~\cite{mukuda,mukuda2}, which were one of the most important ``evidence'' 
for the spin triplet pairing scenario, are in error
due to heating effects by NMR pulses. This result is confirmed by the original researcher~\cite{ishida}.
All recent results unambiguously point to the spin singlet pairing under the strong Pauli effects.
Therefore, it is quite reasonable to expect the LO state to realize in this ``super-clean'' material.
Moreover, its quasi two-dimensional electronic structure is also favorable for it.
Here we examine its possibility in light of the present calculations.

(A) According to the specific heat experiment data~\cite{kittakaPC,yonezawaPC},
$\gamma(H)$ at lower $T$ exhibits an anomaly just before the first order jump at $H^{ab}_{\rm c2}=1.5$T
where almost linear and monotonous $\gamma(H)$ in $H$ deviates upwardly around $H$=1.2T at $T$=0.13K~\cite{yonezawaPC}.
This behavior is similar to Figs.~\ref{fig:thermo1}(d) and ~\ref{fig:thermo1}(h). 
Thus we can identify $H_{\rm LO}\sim 1.2$T at that $T$.

(B) The ultra-high resolution magnetostriction experiment~\cite{striction} 
is performed and detects two successive anomalies as a function of $H$
at low $T$, corresponding to $H_{\rm LO}$ and $H_{\rm c2}$.
The two first order lines $H_{\rm LO}$ and $H_{\rm c2}$ merge at $H_{\rm Lifshitz}$=1.2T and $T_{\rm Lifshitz}$=0.8K
which should be the tricritical Lifshitz point. Thus the constructed phase diagram is consistent 
with our Figs.~\ref{fig:phase-diagram1}(b) and ~\ref{fig:phase-diagram1}(d) qualitatively.
Note that the angle-resolved specific heat measurement~\cite{angle} also detects
the anomalous oscillation sign change at higher $H$ regions, signaling the LO phase.
  
(C) One of the most direct visualizations of the LO state is to use 
STM measurement under parallel fields. 
As shown in Fig.~\ref{fig:profile} (also see Fig. 6 in Ref.~\onlinecite{ichiokaFFLO}),
the nodal plane can be imaged as a distinctive stripe structure near the
zero-bias energy region in STM-STS experiment.
This stripe image is best observed under an applied field parallel to
the surface of the $ab$ plane where the vortices lying near the surface. 
The estimated stripe distance varies, depending on the field strength as seen from
Fig.~\ref{fig:phase-diagram1}(a) and Fig.~\ref{fig:phase-diagram1}(c), typically $L=20\xi\sim 200$nm
with $\xi\sim$10nm.
Since in this STM parallel configuration, the vortex lattices are successfully imaged before
in 2H-NbSe$_2$~\cite{hess,suderow,wei}, this can be a feasible experiment on Sr$_2$RuO$_4$
in which STM experiment is done~\cite{firmo}.

(D)
According to the recent $^{17}$O-NMR experiment~\cite{ishidaPC}, the NMR spectrum is split at around $H$=1.35T
and $T$=0.07K for the in-plane field. This double horn spectrum is akin to our result shown in Fig.~\ref{fig:PM}.
The corresponding $H$-$T$ region also coincides roughly with the LO phase diagram given by Kittaka {\it et al.}~\cite{kittakaPC}.

(E)
The $\bf q$-vector direction of the LO state is anticipated in Sr$_2$RuO$_4$
as follows: There are three bands $\alpha$, $\beta$, and $\gamma$.
The first two have squared cross-sectional shapes in the $ab$ plane while
$\gamma$ Fermi surface is somewhat rounded. The best nesting for the LO phase
is that the $\bf q$-vector points to (110) direction rather than (100) 
because (110) direction nests two-sides of the squared Fermi surface simultaneously
and more advantageous than (100). This can be confirmed by calculating superconducting susceptibility
based on first principles band calculation~\cite{suzukiPC}.
Since the $\bf q$-vector is fixed
to either (110) or (1$\overline{1}$0) under the in-plane $H$, it happens
that when rotating $H$ in the $ab$ plane a switching phenomenon from (110) or (1$\overline{1}$0)
may be observed, similar to that observed in CeCoIn$_5$~\cite{simon,movshovich}.

(F)
The SANS experiments on Sr$_2$RuO$_4$ done so far~\cite{morten214,kuhn}
only probe the transverse component relative to the  field direction
nearly applied to the $ab$ plane.
The Pauli paramagnetic effect manifests itself in the longitudinal component which is
discussed above. Thus the existing data do not provide us the information on the LO state.
In principle, it is possible to perform the SANS experiment to see the longitudinal component.
At present due to the low neutron flux intensity and/or the uniformity of 
the applied magnetic field~\cite{mortenPC} prevent us from observing it.

\subsection{CeCoIn$_5$}
The heavy Fermion superconductor CeCoIn$_5$ is one of the prime candidates for realizing LO state.
Many experimental and theoretical works have been already devoted to studying it in this respect and accumulated
several important clues for LO state. Here in the light of the present theory, we examine its possibility and
propose further experimental and theoretical verifications toward this end.

CeCoIn$_5$ is known for a superconductor with strong Pauli paramagnetic effect
because of the strong $H_{\rm c2}$ suppression~\cite{onuki}, the first order transitions at $H_{\rm c2}$ both
for $H\parallel ab$ and $H\parallel c$ observed by specific heat~\cite{bianchi} and magnetization~\cite{tayama} measurements.
This system is favorable for the LO state since the coherence length is short ($\xi^{ab}$=8.2nm and $\xi^{c}$=3.5nm)
due to heavy effective mass compared to the mean free path $l\sim 1000$nm, thus it is a clean system, 
and the Maki parameter $\mu\sim10$ is large enough.
Thus it is legitimate to seek the LO state in this material.
Since for $H\parallel ab$,  the situation is complicated by the existence of the so-called Q-phase~\cite{kenzelmann,simon},
which is a mixture of the antiferromagnetism and LO state, we mainly focus on the simpler case of  $H\parallel c$.

(A) NMR

\noindent
We start to discuss the NMR experiments on CeCoIn$_5$ \cite{kumagai1,kumagai2}. 
The observed double peak structure of In(2a) of the NMR spectra for $H\parallel c$
and for $H\parallel ab$ is remarkably similar to our Fig.~\ref{fig:PM}(a)
(see the spectral evolutions in Fig. 1 of Ref.~\onlinecite{kumagai1} and Fig. 2 of Ref.~\onlinecite{kumagai2}).
The proposed phase diagram of the LO state for $H\parallel c$ is also 
similar to our Fig.~\ref{fig:phase-diagram1}(b) and Fig.~\ref{fig:phase-diagram1}(d) 
where $H_{\rm LO}/H_{c2}\sim 0.975$ for $\mu=5$
compared with $H_{\rm LO}/H_{c2}=4.7T/4.95T\sim 0.95$ at low temperatures 
for $H\parallel c$~\cite{kumagai1}. 
As mentioned before the value of $H_{\rm LO}/H_{c2}$
depends on $\mu$, but the topological shape of the LO phase diagram
is hardly changed as compared with Fig.~\ref{fig:phase-diagram1}(b) for $\mu=5$ and Fig.~\ref{fig:phase-diagram1}(d) for $\mu=2$. 
In this connection, for $H\parallel ab$ the proposed 
phase diagram (see Fig. 3 of Ref.~\onlinecite{kumagai2}) is quite modified because of the presence of the existing
SDW whose origin is debated. Generally heavy Fermion superconductors have a tendency to the 
SDW instability~\cite{suzuki,kato}. 

We also point out that the observed 
$M_{\rm para}(H)$ (see Fig. 4 of Ref.~\onlinecite{kumagai2}),
which shows a strong rise at the onset of the LO state, is again very similar to our results 
in Fig.~\ref{fig:thermo1}(c) and Fig.~\ref{fig:thermo1}(f).
Therefore, judging from those features: the spectral shape and the field evolution of $M_{\rm para}(H)$, we conclude
that in the high fields for $H\parallel c$ the genuine LO phase is realized in this system.

(B) Entropy and specific heat

\noindent
In order to confirm this identification, we consider other thermodynamic measurements.
Tokiwa {\it et al.}~\cite{tokiwa} measured the specific heat and magneto-caloric effect
and found a kink in the entropy $dS(H)/dH$ at $H\sim4.4$T of $T$=0.2K which coincides with
the expected LO phase diagram. However, the calculated $S(H)$ behaviors shown in Figs.~\ref{fig:thermo1}(c)
and ~\ref{fig:thermo1}(g) are not reproduced precisely. This origin is not known at this moment.

(C) SANS

\noindent
White {\it et al.}~\cite{white} performed the SANS experiment for $H\parallel c$
and studied the vortex lattice structure in this system.
Apart from interesting vortex lattice symmetry changes as a function of $H$,
they observe the fundamental form factor $F_{100}(H)$ (see Fig. 1 in Ref.~\onlinecite{white}),
which is to favorably compared with the insets of Figs.~\ref{fig:SANS1}(b) and ~\ref{fig:SANS1}(f).
Namely, $F_{100}(H)$ gradually increases  and suddenly drops just before $H_{\rm c2}$,
which should be contrasted with the ordinary type II superconductors with the
monotonous and exponential decrease of $F_{100}(H)$.
 
So far, the detailed SANS observation inside the LO phase is not done yet.
There is no data for other form factors to be compared in particular $F_{102}(H)$ of Figs.~\ref{fig:SANS1}(c) and ~\ref{fig:SANS1}(g),
which are a hallmark to the LO state.
Here we point out the feasibility to observe $F_{102}(H)$ in this system.
According to our calculations shown in Fig.~\ref{fig:SANS1},
the anticipated intensity of $|F_{102}|^2$ is one or two order magnitude smaller than $|F_{100}|^2$.
We emphasize that this intensity is already covered by the $|F_{100}|^2$ observation~\cite{white}, meaning that $|F_{102}|^2$
can be detected by the present facility and quite feasible.
Thus we challenge SANS experimentalists to perform it
in order to establish the LO state unambiguously.

(D) STM

\noindent
One of the most difficult tasks for STM experiment is to prepare a high quality surface,
which is not always possible, depending on materials.
CeCoIn$_5$ is fortunate because the STM-STS measurements are already performed~\cite{yazdani,yazdani2}
and guaranteed to prepare a good surface.
Then we propose the same parallel field STM-STS measurement to
observe the nodal stripe structure associated with the LO state discussed earlier.
Since judging from the amplitude of the paramagnetic moment jump at $H_{\rm c2}$
thermodynamic signature of the LO state in CeCoIn$_5$ is far clearer than that in Sr$_2$RuO$_4$,
we understand that CeCoIn$_5$ is the best candidate for confirming the LO state by STM-STS too.

\subsection{CeCu$_2$Si$_2$}

Kitagawa, {\it et al.}~\cite{kitagawa} have performed NMR measurements on CeCu$_2$Si$_2$
and found that $1/TT_1$ as a function of $H$ enhances just near $H_{\rm c2}$.
Since $1/TT_1\propto N(E=0)^2$, this behavior is similar to that of LO phase shown in 
Figs.~\ref{fig:thermo1}(d) and ~\ref{fig:thermo1}(h). This lets the authors claim the evidence for the LO state.
It is true that this system is under strong Pauli paramagnetic effect because of the severe  $H_{\rm c2}$ suppression observed.
However, in view of high residual resistance at lower $T$, meaning that the mean free path is short
and multiband nature, the LO interpretation must be cautious.
In fact, we argue~\cite{CeCu2Si2} that the absence of the first order transition at  $H_{\rm c2}$ in this system
can be understood in terms of the interplay of multi-bands, which hides otherwise the first order transition
expected for a single band. We also point out that the zero-energy DOS $N(E=0)$ can be enhanced more than 
the normal DOS at high $H$, which could explain the enhanced $1/TT_1$ phenomenon.
Indeed this is observed in the specific heat experiment~\cite{kittakaCeCu2Si2}. 
This is consistent with the STM observation~\cite{wahl} too.
Thus we conclude that there is no evidence for LO state in CeCu$_2$Si$_2$.

\subsection{(BEDT-TTF)$_2$X}

The organic superconductors (BEDT-TTF)$_2$X (X=Cu(NCS)$_2$~\cite{lortz,bergk,wright,vesna2,agosta} 
and SF$_5$CH$_2$CF$_2$SO$_3$~\cite{sugiura,uji}) are ideal
candidates for LO state.
The first order phase transition at $H_{\rm c2}$ is observed by the specific heat~\cite{lortz} 
and magnetic torque~\cite{bergk} measurements
in X=Cu(NCS)$_2$.
Agosta {\it et al.}~\cite{agosta}  measured the field dependent specific heat and found a sharp increase of it similar to our Figs.~\ref{fig:thermo1}(d) and
~\ref{fig:thermo1}(h) where at the onset field $H_{\rm LO}$, the phase transition is found to be of first order with hysteresis.
This behavior is also backed up by NMR experiment~\cite{vesna2} where $1/TT_1$ as a function of $H$ enhances just near $H_{\rm c2}$.
The phase diagram obtained~\cite{agosta} with the enhanced $H_{\rm c2}$ and
wider LO region  is somewhat different from those in Figs.~\ref{fig:phase-diagram1}(b)
and ~\ref{fig:phase-diagram1}(d). This difference may come from the different vortex nature in this organic superconductors.
It is the Josephson type vortex without vortex core, and only phase is winding around.
Thus the orbital depairing effect is less severe here, stabilizing the LO at higher fields
compared with our case. Since no one succeeded in microscopically describing the
Josephson vortex nature, it is difficult  to reproduce the LO phase diagram.
The situation may be more akin to the cases without orbital depairing.
In fact according to Machida and Nakanishi~\cite{nakanishi}, the phase diagram 
with diverging $H_{\rm c2}$ is similar to that obtained experimentally~\cite{agosta},
although the divergence itself is an artifact due to quasi-1D band modeling, but the tendency captures
the essential point.

As for X=SF$_5$CH$_2$CF$_2$SO$_3$, the phase diagram is obtained~\cite{sugiura}, which is similar 
to our Figs.~\ref{fig:phase-diagram1}(b) and ~\ref{fig:phase-diagram1}(d),
but the LO region is much wider than ours.
The estimated LO wave length~\cite{uji}
normalized by the coherence length is 2.2$\sim$13.1, which is somewhat shorter 
than our estimate in Fig.~\ref{fig:SANS1}(a) and Fig.~\ref{fig:SANS1}(e).

 \section{Conclusion}

We quantitatively explore the field evolution of the LO states 
for the typical and canonical example of 3D Fermi sphere and $s$-wave pairing, 
by selfconsistently solving the microscopic Eilenberger equation 
in the 3D space of vortex lattice and the LO modulation along the field direction. 
Our calculation, which is reliable in the quantitative level,  fully considers the 
Pauli paramagnetic and orbital depairing effects simultaneously.
In order to facilitate the identification of the LO state by experiments,
we estimate $H$-$T$ phase diagram, NMR spectrum, FLL form factors by SANS, and other thermodynamic
quantities, such as  paramagnetic moment, entropy, and zero-energy density of states
as a function of the magnetic field in FFLO vortex states.
We compare two cases of strong and intermediate Pauli paramagnetic effect. 
We also discuss several candidate materials in the light of the
present theory.

\section*{Acknowledgments}  
We thank T. Sakakibara, S. Kittaka, Y. Shimizu, M. R. Eskildsen, N. Kikugawa, S. Yonezawa, K. Ishida, S. Kitagawa
 for useful discussions on the experimental side, and  N. Nakai, T. Mizushima, M. Takahashi,  Y.  Amano, and M. Ishihara for their collaborations at the earlier stage on this project.
This work is supported by JSPS KAKENHI, No. 17K05553
and partly performed at the Aspen Center for Physics, which is supported by National Science Foundation grant PHY-1607611.

%%%%%%%%%%%%%%%%%%%%%%%%%%%%%%%%%%%%%%%%%%%%%%%%%%%%%%%%%%%%%%%% 

\end{document}